\documentclass[12pt,letterpaper]{article}
\usepackage[utf8]{inputenc}
\usepackage{xcolor}
\usepackage{natbib}
\usepackage[colorlinks=true,linkcolor=blue,anchorcolor=blue,citecolor=blue,filecolor=black,menucolor=black,runcolor=black,urlcolor=blue]{hyperref}
\usepackage{url}
\usepackage{graphicx}
\usepackage{float}

\usepackage[symbol]{footmisc}

\usepackage[margin=1.25in,footskip=0.25in]{geometry}

\usepackage{amsmath, amssymb, mathtools, mathrsfs}

\usepackage{setspace}
\newcommand{\orcid}[1]{\href{https://orcid.org/#1}{\includegraphics[width=10pt]{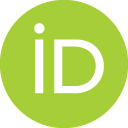}}}

\begin{document}%\thispagestyle{plain}

\section*{\centering \texttt{pocoMC}: A Python package for accelerated Bayesian inference in astronomy and cosmology}

\noindent
\begin{center}
\textbf{Minas Karamanis} \orcid{0000-0001-9489-4612}$^{1}$\footnote[2]{E-mail: minas.karamanis@ed.ac.uk}, \textbf{David Nabergoj} \orcid{0000-0001-6882-627X}$^{2}$, \textbf{Florian Beutler} \orcid{0000-0003-0467-5438}$^{1}$,  \\ \textbf{John A. Peacock} \orcid{0000-0002-1168-8299}$^{1}$, and  \textbf{Uro\v{s} Seljak} \orcid{0000-0003-2262-356X}$^{3}$\\
\textsl{\footnotesize $^{1}$Institute for Astronomy, University of Edinburgh, Royal Observatory, Blackford Hill,\\ Edinburgh EH9 3HJ, UK}\\
\textsl{\footnotesize $^{2}$Faculty of Computer and Information Science, University of Ljubljana, Ve\v{c}na pot 113, 1000 Ljubljana, Slovenia}\\
\textsl{\footnotesize $^{3}$Physics Department, University of California and Lawrence Berkeley National Laboratory Berkeley, CA 94720, USA}
\end{center}

\vspace{0.25cm}
\begin{abstract}
    \texttt{pocoMC} is a Python package for accelerated Bayesian inference in astronomy and cosmology. The code is designed to sample efficiently from posterior distributions with non--trivial geometry, including strong multimodality and non--linearity. To this end, \texttt{pocoMC} relies on the Preconditioned Monte Carlo algorithm which utilises a Normalising Flow in order to decorrelate the parameters of the posterior. It facilitates both tasks of parameter estimation and model comparison, focusing especially on computationally expensive applications. It allows fitting arbitrary models defined as a log--likelihood function and a log--prior probability density function in Python. Compared to popular alternatives (e.g. nested sampling) \texttt{pocoMC} can speed up the sampling procedure by orders of magnitude, cutting down the computational cost substantially. Finally, parallelisation to computing clusters manifests linear scaling. The code is publicly available at \url{https://github.com/minaskar/pocomc}.
\end{abstract}

\thispagestyle{empty}

\section{Statement of need}

Over the past few decades, the volume of astronomical and cosmological data has increased substantially. At the same time, theoretical and phenomenological models in these fields have grown even more complex. As a response to that, a number of methods aiming at efficient Bayesian computation have been developed with the sole task of comparing those models to the available data \citep{trotta2017bayesian, sharma2017markov}. 
In the Bayesian context, scientific inference proceeds though the use of Bayes' theorem:
\begin{equation}\label{eq:bayes}
\mathcal{P}(\theta) = \frac{\mathcal{L}(\theta)\pi(\theta)}{\mathcal{Z}}
\end{equation}
where the posterior $\mathcal{P}(\theta)\equiv p(\theta\vert d,\mathcal{M})$ is the probability of the parameters $\theta$ given the data $d$ and the model $\mathcal{M}$. The other components of this equation are: the likelihood function $\mathcal{L}(\theta)\equiv p(d\vert \theta,\mathcal{M})$, the prior $\pi(\theta) \equiv p(\theta\vert \mathcal{M})$, and the model evidence $\mathcal{Z}=p(d\vert \mathcal{M})$. The prior and the likelihood are usually provided as input in this equation and one seeks to estimate the posterior and the evidence. Knowledge of the posterior, in the form of samples, is paramount for the task of parameter estimation whereas the ratio of model evidences yields the Bayes factor which is the cornerstone of Bayesian model comparison.

Markov chain Monte Carlo (MCMC) has been established as the standard tool for Bayesian computation in astronomy and cosmology, either as a standalone algorithm or as part of another method (e.g. nested sampling \citep{skilling2006nested}). However, as MCMC relies on the local exploration of the posterior, the presence of non-linear correlation between parameters and multimodality can at best hinder its performance and at worst violate its theoretical guarantees of convergence (i.e. ergodicity). Usually, those challenges are partially addressed by reparameterising the model using a common change--of--variables parameter transformation. However, guessing the right kind of reparameterisation a priori is not trivial as it often requires a deep knowledge of the physical model and its symmetries. These problems are usually complicated further by the substantial computational cost of evaluating astronomical and cosmological models. \texttt{pocoMC} is designed to tackle exactly these kinds of difficulties by automatically reparameterising the model such that the parameters of the model are approximately uncorrelated and standard techniques can be applied. As a result, \texttt{pocoMC} produces both samples from the posterior distribution and an unbiased estimate of the model evidence thus facilitating both scientific tasks with excellent efficiency and robustness. Compared to popular alternatives such as nested sampling, \texttt{pocoMC} can reduce the computational cost, and thus, the total run time of the analysis by orders of magnitude, in both artificial and realistic applications \citep{karamanis2022pmc}. Finally, the code is well-tested and is currently used for research work in the field of gravitational wave parameter estimation \citep{vretinaris2022postmerger}.

\begin{figure}[H]
    \centering
	\centerline{\includegraphics[scale=0.175]{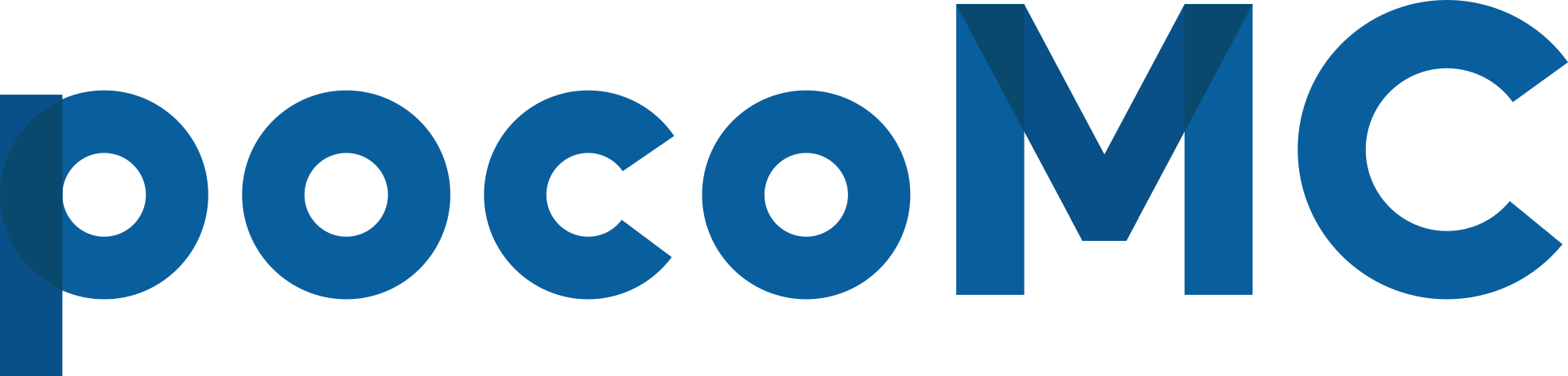}}
    \caption{Logo of \texttt{pocoMC}.}
    \label{fig:logo}
\end{figure}

\section{Method}

\texttt{pocoMC} implements the Preconditioned Monte Carlo (PMC) algorithm. PMC combines the popular Sequential Monte Carlo (SMC) \citep{del2006sequential} method with a Normalising Flow (NF) \citep{papamakarios2021normalizing}. The latter works as a preconditioner for the target distribution of the former. As SMC evolves a population of particles, starting from the prior distribution and gradually approaching the posterior distribution, the NF transforms the parameters of the target distribution such that any correlation between parameters or presence of multimodality is removed. The effect of this bijective transformation is the substantial rise in the sampling efficiency of the algorithm as the particles are allowed to sample freely from the target without being hindered by its locally--curved geometry. The method is explained in detail in the accompanying publication \citep{karamanis2022pmc} and we provide only a short summary here.

\subsection{Sequential Monte Carlo}

The basic idea of basic SMC is to sample from the posterior distribution $\mathcal{P}(\theta)$ by first defining a path of intermediate distributions starting from the prior $\pi(\theta)$. In the
case of \texttt{pocoMC} the path has the form:
\begin{equation}\label{eq:path}
p_{t}(\theta) = \pi(\theta)^{1-\beta_{t}} \mathcal{P}(\theta)^{\beta_{t}}
\end{equation}
where $0=\beta_{1}<\beta_{2}<\dots<\beta_{T}=1$. Starting from the prior, each distribution with density $p_{t}(\theta)$ is sampled in turn using a collection of particles propagated by a number of MCMC steps. Prior to MCMC sampling, the particles are re-weighted using importance sampling and then re-sampled to account for the transition from $p_{t}(\theta)$ to $p_{t+1}(\theta)$. \texttt{pocoMC} utilises the importance weights of this step to define an estimator for the effective sample size (ESS) of the population of particles. Maintaining a fixed value of ESS during the run allows \texttt{pocoMC} to adaptively specify the $\beta_{t}$ schedule.

\subsection{Preconditioned Monte Carlo}

In vanilla SMC, standard MCMC methods (e.g. Metropolis-Hastings) are used to update the positions of the particles during each iteration. This however can become highly inefficient if the distribution $p_{t}(\theta)$ is characterised by a non--trivial geometry. \texttt{pocoMC}, which is based on PMC, utilises a NF to learn an invertible transformation that simplifies the geometry of the distribution by mapping $p_{t}(\theta)$ into a zero-mean unit-variance normal distribution. Sampling then proceeds in the latent space in which correlations are substantially reduced. The positions of the particles are transformed back to the original parameter space at the end of each iteration. This way, PMC and \texttt{pocoMC} are able to sample from very challenging posteriors very efficiently using simple Metropolis-Hastings updates in the preconditioned/uncorrelated latent space.

\section{Features}

\begin{itemize}
    \item User--friendly black-box API (only the log-likelihood, log-prior and some prior samples required from the user)
    \item Default configuration sufficient for most applications (no tuning is required but is possible for experienced users)
    \item Posterior corner, trace, and run plotting tools
    \item Support for both MAF and RealNVP normalising flows with added regularisation \citep{papamakarios2017masked, dinh2016density}
    \item Straightforward parallelisation using MPI or multiprocessing
    \item Continuous integration, unit tests and wide range of examples available
    \item Extensive documentation available  online \url{http://pocomc.readthedocs.io}
\end{itemize}

\section*{Acknowledgements}

MK would like to thank Jamie Donald-McCann and Richard Grumitt for providing constructive comments and George Vretinaris for feedback on an early version of the code. This project has received funding from the European Research Council (ERC) under the European Union's Horizon 2020 research and innovation program (grant agreement 853291), and  by the U.S. Department of Energy, Office of Science, Office of Advanced Scientific Computing Research under Contract No. DE-AC02-05CH11231 at Lawrence Berkeley National Laboratory to enable research for Data-intensive Machine Learning and Analysis. FB is a University Research Fellow.

% end matter

%\bibliographystyle{unsrt}
\bibliographystyle{style}
\bibliography{ref.bib}

\begin{thebibliography}{}

\bibitem[\protect\citename{Del~Moral {\em et~al.}, }2006]{del2006sequential}
Del~Moral, Pierre, Doucet, Arnaud, \& Jasra, Ajay. 2006.
\newblock Sequential monte carlo samplers.
\newblock {\em Journal of the Royal Statistical Society: Series B (Statistical
  Methodology)}, {\bf 68}(3), 411--436.

\bibitem[\protect\citename{Dinh {\em et~al.}, }2016]{dinh2016density}
Dinh, Laurent, Sohl-Dickstein, Jascha, \& Bengio, Samy. 2016.
\newblock Density estimation using real nvp.
\newblock {\em arXiv preprint arXiv:1605.08803}.

\bibitem[\protect\citename{Karamanis {\em et~al.}, }2022]{karamanis2022pmc}
Karamanis, Minas, Beutler, Florian, Peacock, John~A, Nabergoj, David, \&
  Seljak, Uro\v{s}. 2022.
\newblock Accelerating astronomical and cosmological inference with
  Preconditioned Monte Carlo.
\newblock {\em in prep}.

\bibitem[\protect\citename{Papamakarios {\em et~al.},
  }2017]{papamakarios2017masked}
Papamakarios, George, Pavlakou, Theo, \& Murray, Iain. 2017.
\newblock Masked autoregressive flow for density estimation.
\newblock {\em Advances in neural information processing systems}, {\bf 30}.

\bibitem[\protect\citename{Papamakarios {\em et~al.},
  }2021]{papamakarios2021normalizing}
Papamakarios, George, Nalisnick, Eric~T, Rezende, Danilo~Jimenez, Mohamed,
  Shakir, \& Lakshminarayanan, Balaji. 2021.
\newblock Normalizing Flows for Probabilistic Modeling and Inference.
\newblock {\em J. Mach. Learn. Res.}, {\bf 22}(57), 1--64.

\bibitem[\protect\citename{Sharma, }2017]{sharma2017markov}
Sharma, Sanjib. 2017.
\newblock Markov chain Monte Carlo methods for Bayesian data analysis in
  astronomy.
\newblock {\em arXiv preprint arXiv:1706.01629}.

\bibitem[\protect\citename{Skilling, }2006]{skilling2006nested}
Skilling, John. 2006.
\newblock Nested sampling for general Bayesian computation.
\newblock {\em Bayesian analysis}, {\bf 1}(4), 833--859.

\bibitem[\protect\citename{Trotta, }2017]{trotta2017bayesian}
Trotta, Roberto. 2017.
\newblock Bayesian methods in cosmology.
\newblock {\em arXiv preprint arXiv:1701.01467}.

\bibitem[\protect\citename{Vretinaris {\em et~al.},
  }2022]{vretinaris2022postmerger}
Vretinaris, George, Vretinaris, Stamatis, Mermigkas, Christos, Karamanis,
  Minas, \& Stergioulas, Nikolaos. 2022.
\newblock Robust and fast parameter estimation of gravitational waves from
  neutron star merger remnants.
\newblock {\em in prep}.

\end{thebibliography}

%\appendix

\end{document}